\documentclass[slac_one]{revtex4}
\usepackage{graphicx}
\usepackage{fancyhdr}
\begin{document}

\title{Future Antiproton Experiments at Fermilab} 

%

\author{D. M. Kaplan}
\affiliation{Illinois Institute of Technology, Chicago, IL 60616, USA}
\begin{abstract}
Fermilab operates the world's most intense antiproton source. Newly proposed experiments can use those antiprotons either parasitically during Tevatron Collider running or after the Tevatron Collider finishes in about 2010. In particular, the annihilation of 8 GeV antiprotons might make the world's most intense source of tagged $D^0$ mesons, and thus the best near-term opportunity to study charm mixing and, via {\em CP} violation, to search for new physics; a Penning trap and atom interferometer could be used to measure for the first time the gravitational force on antimatter. 
\end{abstract}

\maketitle

\thispagestyle{fancy}


\section{INTRODUCTION} 
Low- and medium-energy antiproton experiments have fruitfully addressed a variety of topics over many years, starting at LEAR and continuing with the Fermilab Antiproton Source and CERN AD. Techniques and energies used in these experiments have ranged from antiproton annihilation at rest up to 8\,GeV, as well as experiments using trapped antiprotons. Physics issues have included the search for glueballs and hybrid mesons, precision studies of hyperon decay and charmonium spectroscopy, and  tests of {\em CP} and {\em CPT} symmetry. Starting in about 2015, the FAIR project~\cite{FAIR} at GSI-Darmstadt will add to this list studies of strange matter, charm, and nuclei far from stability~\cite{FLAIR,PANDA}. 

Table~\ref{tab-sens-comp} compares available antiproton intensities at CERN, Fermilab, and FAIR. Because the Fermilab Antiproton Source uses 120\,GeV protons on target and accumulates at 8\,GeV, it has a significant rate advantage with respect to FAIR. It also can potentially operate full-time, while at FAIR, the PANDA antiproton experiment~\cite{PANDA} will have to share time with other modes of operation at FAIR. This intensity advantage could be maximized by building a new, small storage ring at Fermilab, in which fixed-target collisions would then take place, to allow the Accumulator to stack antiprotons full-time; in this way a $\overline{p}p$ luminosity of $\sim10^{33}$\,cm$^{-2}$s$^{-1}$ could be supported. But even without an accelerator upgrade, operation at  ${\cal L}\approx10^{33}$\,cm$^{-2}$s$^{-1}$ would be possible with 50\% duty factor, and ${\cal L}\approx2\times10^{32}$\,cm$^{-2}$s$^{-1}$ could be achieved with 85\% duty factor, and an upgrade of the Fermilab E835 detector.

\begin{table}[htb]
\begin{center}
\caption{Antiproton Intensities at Existing and Future Facilities}\label{tab-sens-comp}
\begin{tabular}{lcccc}
\hline & \multicolumn{2}{c}{\textbf{Stacking:}} &\textbf{Clock Hours} & $\mathbf{\overline{p}/}$\textbf{Yr}\\
\raisebox{1.5ex}[0pt]{\textbf{Facility}} & \textbf{Rate} $\mathbf{(10^{10}/}$\textbf{hr)}  & \textbf{ Duty Factor} & \textbf{/Yr} & $\mathbf{(10^{13})}$ 
\\
\hline\hline
CERN AD &  &  & 3800 &0.4 \\
 FNAL  (Accumulator) &  20 & 15\% & 5550 &17 \\
FNAL  (New Ring) & 20 & 90\%& 5550 & 100 \\
FAIR ($\stackrel{>}{_\sim}2015$) &  3.5 & 90\% & 2780 & 9\\
\hline
\end{tabular}
\end{center}
\vspace{-.25in}
\end{table}

\section{PROPOSED ANTIPROTON EXPERIMENTS AT FERMILAB}

\subsection{Medium-Energy $\mathbf{\overline p}\mathbf p$-Annihilation Experiment}

By adding a small magnet, tracking and vertex detectors, and TOF counters to the E835 calorimeter, plus modern, high-bandwidth triggering and data-acquisition systems, several important topics can be studied. 

\subsubsection{Charm Mixing and {\em CP} Violation}

After a more than 20-year search, $D^0$--${\overline D}{}^0$ mixing is now established  at 9.2 standard deviations~\cite{HFAG}, thanks mainly  to the $B$ factories. The level of mixing is consistent with the wide range of Standard Model predictions~\cite{Bigi-Uraltsev-Petrov}; however, this does not preclude a significant and potentially detectable contribution from new physics. Since some new-physics models predict different effects in the charge-2/3 (``up-type") quark sector than in the down-type, it is important to carry out such  studies of charm mesons\,---\,the only up-type system for which  meson mixing can occur.

The $\overline{p}p$ annihilation cross section to open charm is expected to be  substantial; for example, a recent estimate gives $\sigma(\overline{p}p\to D^{*0}{\overline D}{}^0)\approx1.3\,\mu$b at $\sqrt{s}=4.2$\,GeV~\cite{Braaten}. At ${\cal L}=2\times10^{32}$\,cm$^{-2}$s$^{-1}$, this represents some $5\times10^9$ events per year, substantially exceeding each year the integrated sample ($10^9$ events) 
available at the $B$ factories. Since there will  also be $D^{*\pm}D^\mp$, $D^*\overline{D}{}^*$, and $D\overline{D}$ events,  the total charm sample will be even larger; with the use of a target nucleus heavier than hydrogen, the charm-production $A$-dependence~\cite{A-dep} could further enhance statistics. Such a target could also localize primary interactions to an ${\cal O}(\mu$m)-sized region, allowing the $D$-meson decay distance to be cleanly determined. Medium-energy ($p_{\overline{p}}\approx8\,$GeV/$c$) $\overline{p}N$ annihilation may thus be the optimal way to study charm mixing and search for possible new-physics contributions via the clean signature~\cite{Petrov} of charm {\em CP} violation (CPV).

\subsubsection{Hyperon {\em CP} Violation and Rare Decays}

The Fermilab HyperCP Experiment~\cite{E871} amassed the world's largest samples of hyperon decays, including $2.5\times10^9$ reconstructed ${}^{^(}\overline{\Xi}{}^{^{\,)\!}}{}^\mp$ decays and $10^{10}$ produced $\Sigma^+$. HyperCP observed unexpected signals at the $\stackrel{>}{_\sim}$\,2$\sigma$ level  for possible new physics in the rare hyperon decay $\Sigma^+\to p\mu^+\mu^-$~\cite{Park-etal} and the  ${}^{^(}\overline{\Xi}{}^{^{\,)\!}}{}^\mp\to{}^{^(}\overline{\Lambda}{}^{^{\,)}}\pi^\mp\to 
{}^{^{(}}\overline{p}{}^{^{\,)}} 
\pi^\mp\pi^\mp$ {\em CP} asymmetry~\cite{BEACH08}. Since the $\overline{p}p\to\Omega^-{\overline\Omega}{}^+$ threshold lies in the same region as the open-charm threshold, the proposed  experiment  can further test these observations using ${}^{^(}\overline{\Omega}{}^{^{\,)\!}}{}^\mp\to{}^{^(}\overline{\Xi}{}^{^{\,)\!}}{}^\mp\mu^+\mu^-$ decay and potential ${}^{^(}\overline{\Omega}{}^{^{\,)\!}}{}^\mp$ CPV~\cite{OmegaCP}. Moreover, the dedicated $\overline{p}$ storage ring of Table~\ref{tab-sens-comp} could decelerate antiprotons to the $\Lambda\overline{\Lambda}$, $\Sigma^+\overline{\Sigma}{}^-$, and $\Xi^-\overline{\Xi}{}^+$ thresholds, where an experiment at $10^{33}$ luminosity could amass the clean, $>10^{10}$-event samples needed to confront the HyperCP signals directly.

\subsubsection{Precision Charmonium Measurements}

Using the Fermilab Antiproton Source,  experiments E760 and E835 made the world's most precise measurements of charmonium masses and widths~\cite{E835}. This precision ($\stackrel{<}{_\sim}$\,100\,keV) was enabled by the  small energy spread of the stochastically cooled antiproton beam and the absence of Fermi motion and negligible energy loss in the H$_2$ cluster-jet target. The other key advantage of  ${\overline p}p$ annihilation is its ability to produce charmonium states of all quantum numbers, in contrast to $e^+e^-$ machines, which produce primarily $1^{--}$ states (and, to a lesser degree, states accessible in $B$ decay or in $2\gamma$ production).
Although charmonium has by now been extensively studied, a number of questions remain, including the nature of the mysterious $X(3872)$ state and improved measurement  of $h_c$ and $\eta^\prime_c$ parameters. 

\subsection{Antihydrogen Experiments} 
\subsubsection{Antihydrogen-in-Flight {\em CPT} Tests}

The study of antihydrogen atoms in flight may be a way around some of the difficulties encountered in the CERN trapping experiments. First steps in this direction were taken by Fermilab E862,  which observed formation of antihydrogen in flight  during 1996--7~\cite{Blanford}. Methods to measure the antihydrogen Lamb shift and fine structure (the $2s_{1/2}$--$2p_{1/2}$ and $2p_{1/2}$--$2p_{3/2}$ energy differences) were subsequently worked out~\cite{Blanford-Lamb-shift}. Progress toward this goal may be compatible with normal Tevatron Collider operations\,---\,a possibility  currently under investigation. If the feasibility of the approach is borne out by further work, the program could continue into the post-Tevatron era.

\subsubsection{Antimatter Gravity Experiment}

While General Relativity predicts that the gravitational forces on matter and antimatter should be identical, no direct experimental test of this prediction has yet been made~\cite{Fischler-etal}. Attempts at a quantized theory of gravity generally introduce non-tensor forces, which could cancel for matter-matter and antimatter-antimatter interactions but add for matter-antimatter ones. In addition, possible ``fifth forces" or non-$1/r^2$ dependence have been discussed. Such effects can be sensitively sought by measuring the gravitational acceleration of antimatter in the field of the earth. While various such experiments have been discussed for many years, one\,---\,measurement of the gravitational acceleration of antihydrogen\,---\,has only recently become feasible and is now proposed both at CERN and at Fermilab~\cite{AEGIS,LoI}.

The principle of the Antimatter Gravity Experiment (AGE) is to form a beam of slow ($\approx1\,$km/s)  antihydrogen in a Penning trap and pass the beam through a $\approx1$-m-long Mach-Zehnder interferometer. The phase of the interference pattern can be measured to a small fraction of the ($\approx1\,\mu$m) grating period, and measurement of the phase vs.\ the speed of the atom determines $\bar g$. Simulation shows that $\bar g$ can be measured to 1\% of $g$ with one-million antihydrogen ($\overline H$) atoms incident on the interferometer; this can be done parasitically during the Tevatron run. The proposed AGE goal is a $10^{-4}$ measurement, requiring $10^{10}$ $\overline H$ atoms. Given the expected ${\cal O}(10^{-5})$ antiproton-trapping and $\overline H$-formation efficiency, this can be accomplished at the Antiproton Source in a few-month dedicated run.

\section{OUTLOOK}

With the end of the Tevatron Collider program in sight, new and unique measurements are possible at the Fermilab Antiproton Source~\cite{New-pbar}. Such a program can substantially
 broaden the clientele and appeal of US particle physics.

\begin{acknowledgments}
The author thanks his pbar collaborators,\footnote{Arizona: A.~Cronin; Riverside: A.~P.~Mills, Jr.; Cassino: G.~M.~Piacentino; Duke: T.~J.~Phillips; FNAL: G.~Apollinari, D.~R.~Broemmelsiek, B.~C.~Brown, C.~N.~Brown, D.~C.~Christian, P.~Derwent, M.~Fischler, K.~Gollwitzer, A.~Hahn, V.~Papadimitriou, J.~Volk, S.~Werkema, H.~B.~White, G.~P.~Yeh; Ferrara: W.~Baldini, G.~Stancari, M.~Stancari; Hbar Tech.: J.~R.~Babcock, S.~D.~Howe, G.~P.~Jackson, J.~M.~Zlotnicki; IIT: D.~M.~Kaplan, T.~J.~Roberts, H.~A.~Rubin, Y.~Torun, C.~G.~White; KSU: G.~A.~Horton-Smith, B. Ratra; Luther Coll.: T.~K.~Pedlar; Michigan: H.~R.~Gustafson; Northwestern: J.~Rosen; SMU: T.~Coan; Virginia: E.~C.~Dukes.
}
and E. Braaten for useful conversations. Work supported by Department of Energy grant DE-FG02-94ER40840.
\end{acknowledgments}

\end{document}